\documentclass{article}               
\usepackage[utf8]{inputenc}
\usepackage[T1]{fontenc}
\usepackage[top=1.5cm,left=2cm,right=2cm,bottom=2.5cm]{geometry}
\usepackage{graphicx}
\usepackage{authblk}                  
\usepackage{microtype}   
\usepackage{graphicx}
\usepackage{stfloats}
\usepackage{hyperref}
\usepackage{dirtree}
\usepackage{tcolorbox}
\usepackage{booktabs}

\renewcommand{\thefootnote}{\fnsymbol{footnote}}

\title{\vspace*{-1cm}ReXGroundingCT: A 3D Chest CT Dataset for Segmentation of Findings from Free-Text Reports}

\author[1]{Mohammed Baharoon, BS$^{*,}$}
\author[1]{Luyang Luo, PhD$^{*,}$}
\author[2,3]{Michael Moritz, MD}
\author[4]{Abhinav Kumar, MS}
\author[1,5]{Sung Eun Kim, MD}
\author[1]{Xiaoman Zhang, PhD}
\author[6]{Miao Zhu, MD}
\author[7,8]{Mahmoud Hussain Alabbad, MD}
\author[7,8]{Maha Sbayel Alhazmi, MD}
\author[9]{Neel P. Mistry, MD}
\author[1,10]{Lucas Bijnens, MD}
\author[3]{Kent Ryan Kleinschmidt, BS}
\author[3]{Brady Chrisler, BS}
\author[3]{Sathvik Suryadevara, BS}
\author[3]{Sri Sai Dinesh Jaliparthi, BS}
\author[3]{Noah Michael Prudlo, BS}
\author[3]{Mark David Marino, BS}
\author[3]{Jeremy Palacio, BS}
\author[3]{Rithvik Akula, BS}
\author[11]{Di Zhou, MD}
\author[1]{Hong-Yu Zhou, PhD}
\author[12]{Ibrahim Ethem Hamamci, MD}
\author[9]{Scott J. Adams, MD, PhD}
\author[7,8]{Hassan Rayhan AlOmaish, MD}
\author[1]{Pranav Rajpurkar, PhD}

\affil[1]{Department of Biomedical Informatics, Harvard Medical School, Boston, MA}
\affil[2]{SSM Health, St.\ Louis, MO}
\affil[3]{Saint Louis University School of Medicine, St.\ Louis, MO}
\affil[4]{Icahn School of Medicine at Mount Sinai, New York, NY}
\affil[5]{National Strategic Technology Research Institute, Seoul National University Hospital, South Korea}
\affil[6]{Brigham and Women’s Hospital, Boston, MA}
\affil[7]{Chest Radiology Division, Medical Imaging Department, King Abdullah Specialized Children's Hospital, Riyadh, Saudi Arabia}
\affil[8]{King Abdulaziz Medical City, Ministry of National Guard Health Affairs, Riyadh, Saudi Arabia}
\affil[9]{Department of Medical Imaging, Royal University Hospital, Saskatoon, SK, Canada}
\affil[10]{KU Leuven, Leuven, Belgium}
\affil[11]{Department of Radiology, Eye \& ENT Hospital of Fudan University, Shanghai, China}
\affil[12]{University of Zurich, Zurich, Switzerland}

\date{}

\begin{document}
\maketitle
\begingroup
\renewcommand\thefootnote{*}
\footnotetext{Equal contribution.}
\endgroup
\footnotetext{Contact Email: MohammedSalimAB@outlook.com}

\begin{abstract}

\textbf{BACKGROUND} Connecting free-text descriptions such as “3 mm nodule in the lower left lobe” to precise 3D segmentations remains an unsolved challenge in medical AI. Existing chest CT datasets rely on structured labels or predefined categories, limiting their ability to represent the richness of clinical language and support grounded radiology report generation systems. Bridging this gap requires datasets that capture the expressiveness of free-text radiology findings and link them to anatomically accurate spatial annotations in volumetric imaging.

\textbf{METHODS} We introduce ReXGroundingCT, the first publicly available dataset that links free-text findings to pixel-level 3D segmentations in chest CT scans. The dataset includes 3,142 non-contrast chest CT scans paired with standardized radiology reports from CT-RATE. Construction followed a structured three-stage pipeline. First, GPT-4 was used to extract and standardize findings, descriptors, and metadata from radiology reports originally written in Turkish and machine-translated into English. Second, GPT-4o-mini categorized each finding into a hierarchical ontology of lung and pleural abnormalities. Third, 3D annotations were produced on all CT volumes: the training set was annotated by  quality assured by board-certified radiologists, while the validation and test sets were labeled completely by board-certified radiologists. Additionally, we constructed a complementary chain-of-thought dataset that provides step-by-step hierarchical anatomical reasoning for localizing findings within the CT volume, using GPT-4o and guided by localization coordinates derived from organ segmentation models.

\textbf{RESULTS} ReXGroundingCT provides 16,301 annotated entities across 8,028 text-to-3D-segmentation pairs spanning diverse radiological patterns from 3,142 non-contrast CT scans. Approximately 79\% of findings are focal abnormalities, while 21\% are non-focal. The dataset includes a public validation set of 50 cases and a private test set of 100 cases, both annotated by board-certified radiologists. Model performance on the private test set is hosted on a public leaderboard at \href{https://rexrank.ai/ReXGroundingCT}{https://rexrank.ai/ReXGroundingCT}.

\textbf{CONCLUSIONS} ReXGroundingCT is the first manually curated dataset linking free-text chest CT findings to 3D segmentation masks, establishing a valuable benchmark for developing and evaluating free-text medical segmentation models. It sets the foundation for enabling free-text finding segmentation and grounded radiology report generation in CT imaging. The dataset can be accessed at \href{https://huggingface.co/datasets/rajpurkarlab/ReXGroundingCT}{https://huggingface.co/datasets/rajpurkarlab/ReXGroundingCT.}

\end{abstract}

\twocolumn

\begin{figure*}[t]
    \centering
    \includegraphics[width=1\textwidth]{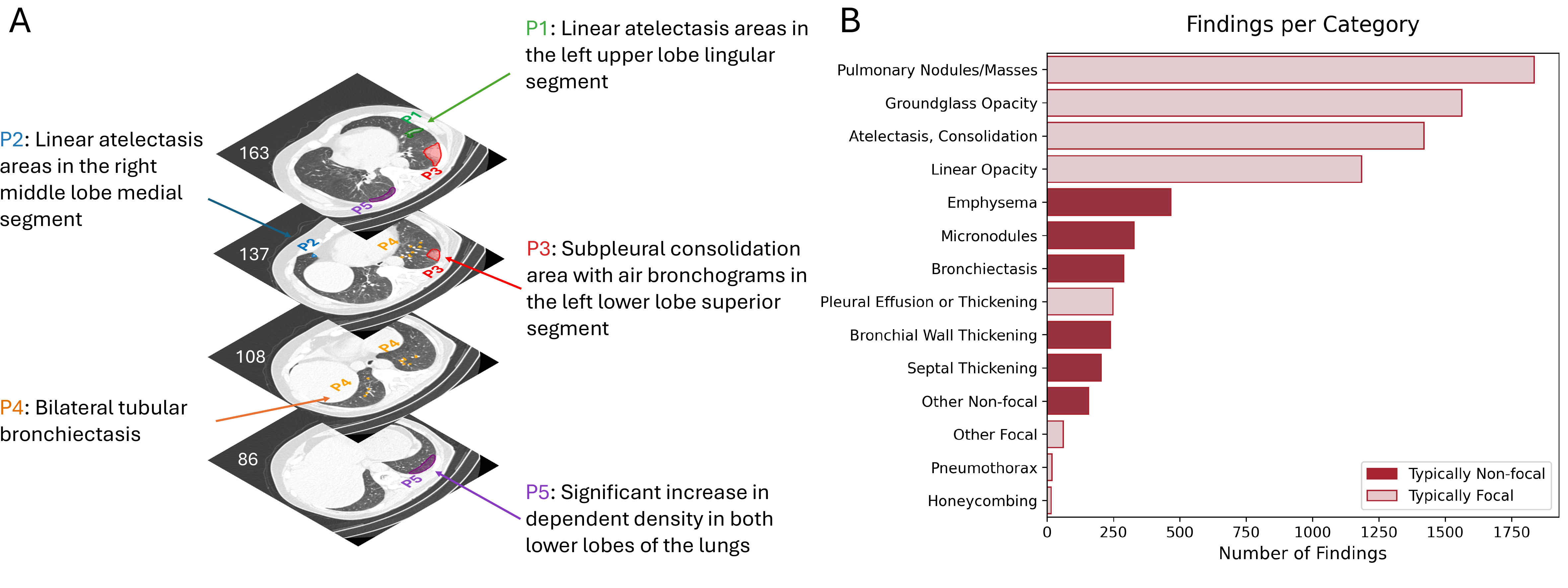}
    \caption{(A) Examples of mapping narrative free-text findings from a radiology report to precise 3D spatial locations in a chest CT scan. Five representative findings illustrate various anatomical sites and patterns of disease. Numbers indicate CT slice locations where the findings are visualized. (B) Distribution of findings per category across the dataset, grouped by typical pattern (non-focal vs. focal).}
    \label{fig:annotation_example}
\end{figure*}

\section{Introduction}
Artificial intelligence (AI) is increasingly transforming healthcare, driving the development of systems that can interpret complex medical images and assist clinical decision-making \cite{silcox2024potential}. However, deployment in clinical practice depends not only on technical performance, but also on generating outputs that are explainable and verifiable \cite{sadeghi2024review, abgrall2024should}. One emerging frontier is grounded radiology report generation (GRRG), which aims to produce detailed textual descriptions of imaging findings paired with precise spatial localizations in the imaging volume, linking language and anatomy in a clinically meaningful way \cite{bannur2024maira, de2025padchest, rao2025multimodal, fahrner2025generative}. 

A central challenge for GRRG lies in bridging the gap between unstructured, free‑text radiology reports and precise three‑dimensional localization of findings within chest CT scans. Translating these narrative descriptions into spatially grounded labels remains largely unsolved. Existing chest CT datasets provide categorical labels  or lesion masks but cannot represent the rich variability of free‑text language \cite{de2025uls23, wasserthal2023totalsegmentator, armato2011lung}.

We introduce \textbf{ReXGroundingCT}, the first publicly available manually annotated dataset connecting free‑text radiology findings to precise 3D spatial localizations in chest CT imaging. Unlike previous data sets focused on predefined disease categories, ReXGroundingCT captures a diverse descriptive language and pairs it with detailed 3D segmentation masks. The dataset was developed via a multi‑stage pipeline involving report standardization, finding extraction and categorization, and pixel‑level annotation of CT scans by trained annotators and board‑certified radiologists. An example of a data point from ReXGroundingCT is shown in Figure \ref{fig:annotation_example}, A. 

\section{Methods}

\subsection{Source Data}

ReXGroundingCT is built upon CT-RATE, a publicly available dataset pairing non-contrast 3D chest CT volumes with corresponding radiology reports \cite{hamamci2024developing}. CT-RATE includes 25,692 chest CT studies, reconstructed into 50,188 volumes, totaling over 14.3 million axial slices, acquired between May 2015 and January 2023 at Istanbul Medipol University Mega Hospital in Turkey. The dataset spans 21,304 unique patients with an average age of 48.8 years and a sex distribution of 41.6\% female and 58.4\% male.

Radiology reports in CT-RATE are structured into four sections: clinical information, technique, findings, and impression. Originally written in Turkish, reports were machine-translated into English and corrected by bilingual medical students for clarity and consistency. Extensive anonymization was performed to remove personal identifiers from both the imaging data and the reports.

For ReXGroundingCT, we curated 3,142 chest CT volumes from CT-RATE, selected sequentially in groups by increasing CT-RATE ID. Because IDs are randomly assigned, this yields a representative random sample. Each selected volume retains its English report and metadata, including patient demographics and scan parameters. 

\subsection{Report Rewriting}

A qualitative review by a board-certified radiologist found that reports in the CT-RATE dataset often contained non-standard terminology and inconsistent phrasing due to translation artifacts and varied reporting styles. To address this, we developed a report rewriting pipeline using GPT-4 \cite{achiam2023gpt} to produce standardized reports with terminology and phrasing aligned with conventions typical of U.S. radiology practice.

The model was instructed to correct non-standard terminology to align with conventions typical of American radiology practice and to improve grammatical clarity, while preserving all clinical details, measurements, and anatomical references. For example, terms such as “millimetric” were rewritten as “subcentimeter,” “pleuroparenchymal sequelae” was simplified to “scarring” or “fibrosis,” and phrases like “no space-occupying lesion” were replaced with “no lesion.” Irrelevant statements that were not related to clinical interpretation were removed to improve the conciseness and clarity of the report.

Quality control was performed through manual review by a board-certified radiologist. Over 50 reports were qualitatively reviewed and rated on a binary acceptable versus not acceptable basis, with all 50 reports deemed acceptable by the radiologist. The rewriting prompt used in this process was developed through a multistep design in collaboration with the radiologist to ensure high fidelity to clinical meaning and reporting standards. These standardized reports formed the basis for subsequent abnormality extraction, categorization, and annotation in ReXGroundingCT.

\subsection{Abnormality Extraction and Categorization}

Following the rewriting of CT-RATE reports, we implemented a two-stage pipeline to extract and categorize abnormalities from each standardized radiology report.

For abnormality extraction, we prompted GPT-4 to systematically analyze the rewritten Findings and Impressions sections, isolating distinct anatomical observations, whether normal or abnormal. The model was instructed to split sentences describing multiple findings into separate entries and to consolidate information spread across multiple sentences into unified, detailed findings. Extracted phrases included details such as lesion size, morphology, anatomical location, and any references to prior imaging studies indicating stability or change over time. The extraction process explicitly excluded interpretive language, diagnostic speculation, and clinical recommendations, focusing solely on observable imaging features. Measurements were normalized to a consistent format, rounding millimeter values to whole numbers and limiting centimeter values to one decimal place. The model’s output listed each extracted finding alongside its abnormality status and any indication of reference to prior imaging. 

After extraction, each finding was categorized using GPT-4o-mini into a two-level hierarchical schema comprising 12 parent categories and 61 subcategories, shown in Figure \ref{fig:finding_categories}, developed in consultation with a board-certified radiologist to reflect a comprehensive spectrum of chest CT findings encountered in clinical practice. This schema accommodates both focal abnormalities, such as nodules or localized consolidations, and non-focal patterns, including diffuse interstitial changes and widespread emphysema, as well as a broad range of non-pulmonary findings. Categorization was performed using GPT-4o-mini, which received the text of each extracted phrase and assigned it to the appropriate subcategory.

\begin{figure*}[t!]
    \centering
    \includegraphics[width=1\textwidth]{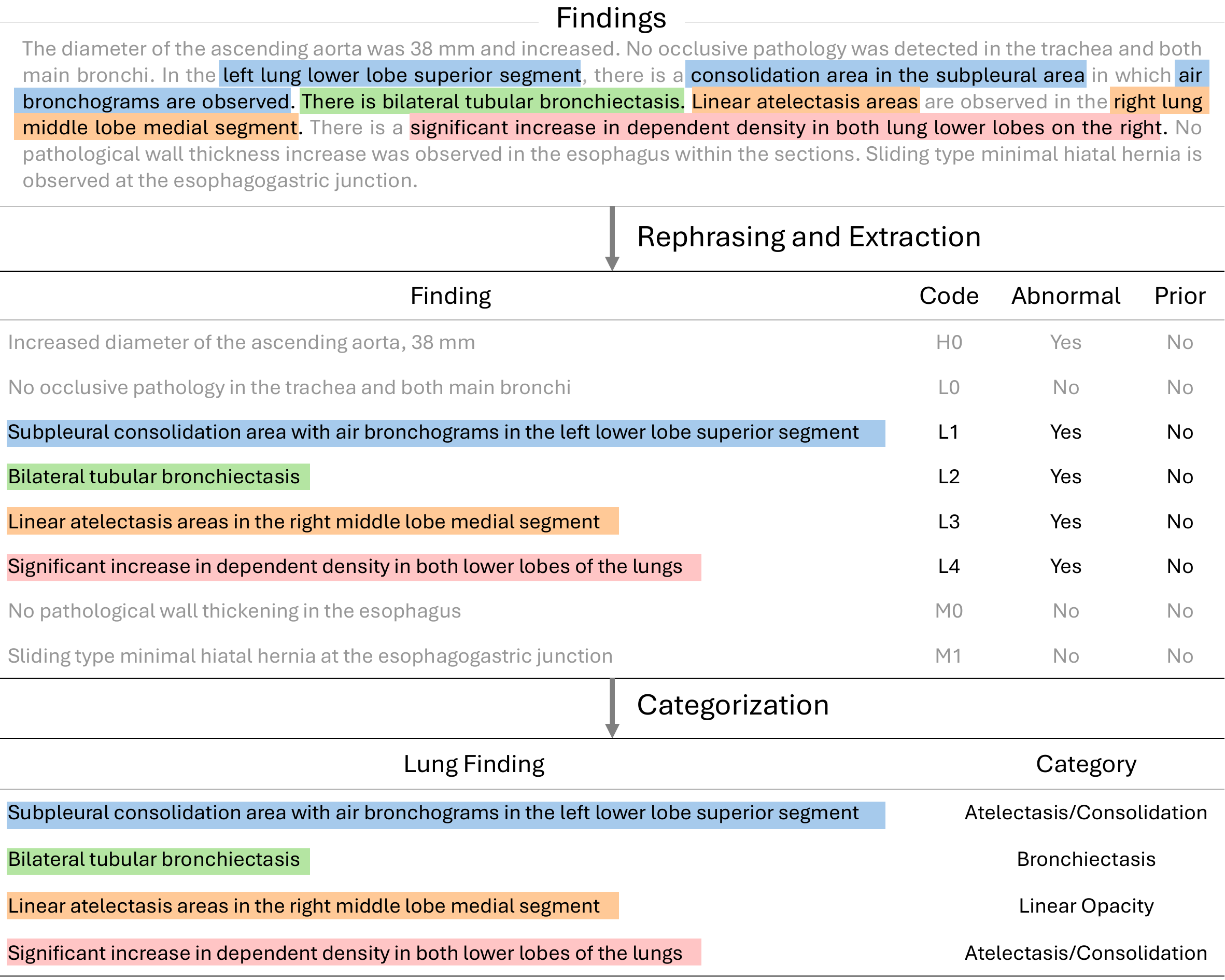}
    \caption{Illustration of the structured pipeline for converting narrative radiology reports into standardized reports. The example shows how a free-text report is parsed into discrete findings, each assessed for abnormality and prior reference, followed by categorization into defined ontology classes. Findings that have a code that starts with "H" refer to "Heart/Vessels" findings, where "L" refer to "Lungs/Airways/Pleura" findings, and "M" refers to "Mediastinum/Hila".}
    \label{fig:report_pipeline}
\end{figure*}

To evaluate the accuracy of the extraction and categorization pipeline, we conducted a manual review of 100 randomly sampled reports, assessed by a board-certified radiologist. Each extracted finding was evaluated for (1) missing descriptors—defined as omissions of essential details such as anatomical location, shape, or measurement—(2) false positives, and (3) false negatives. No false positive extractions were observed, and the rates of false negatives were low: 0.05 in the rephrasing step and 0.01 in the extraction step. Descriptor omissions occurred at rates of 0.14 in the rephrasing step and 0.13 in the extraction step, indicating strong fidelity between the extracted findings and the original report content. For categorization, GPT-4o-mini achieved a precision and recall of 0.99 for lung and pleural findings across the full 12-category ontology. In a subset of 100 findings from categories 1 and 2 (Figure~\ref{fig:finding_categories}), GPT-4o-mini attained an F1 score of 0.92 for subcategory classification. An example of report extraction and categorization pipeline is illustrated in Figure~\ref{fig:report_pipeline}.

\subsection{Annotation}

\begin{figure*}[t!]
    \centering
    \includegraphics[width=1\textwidth]{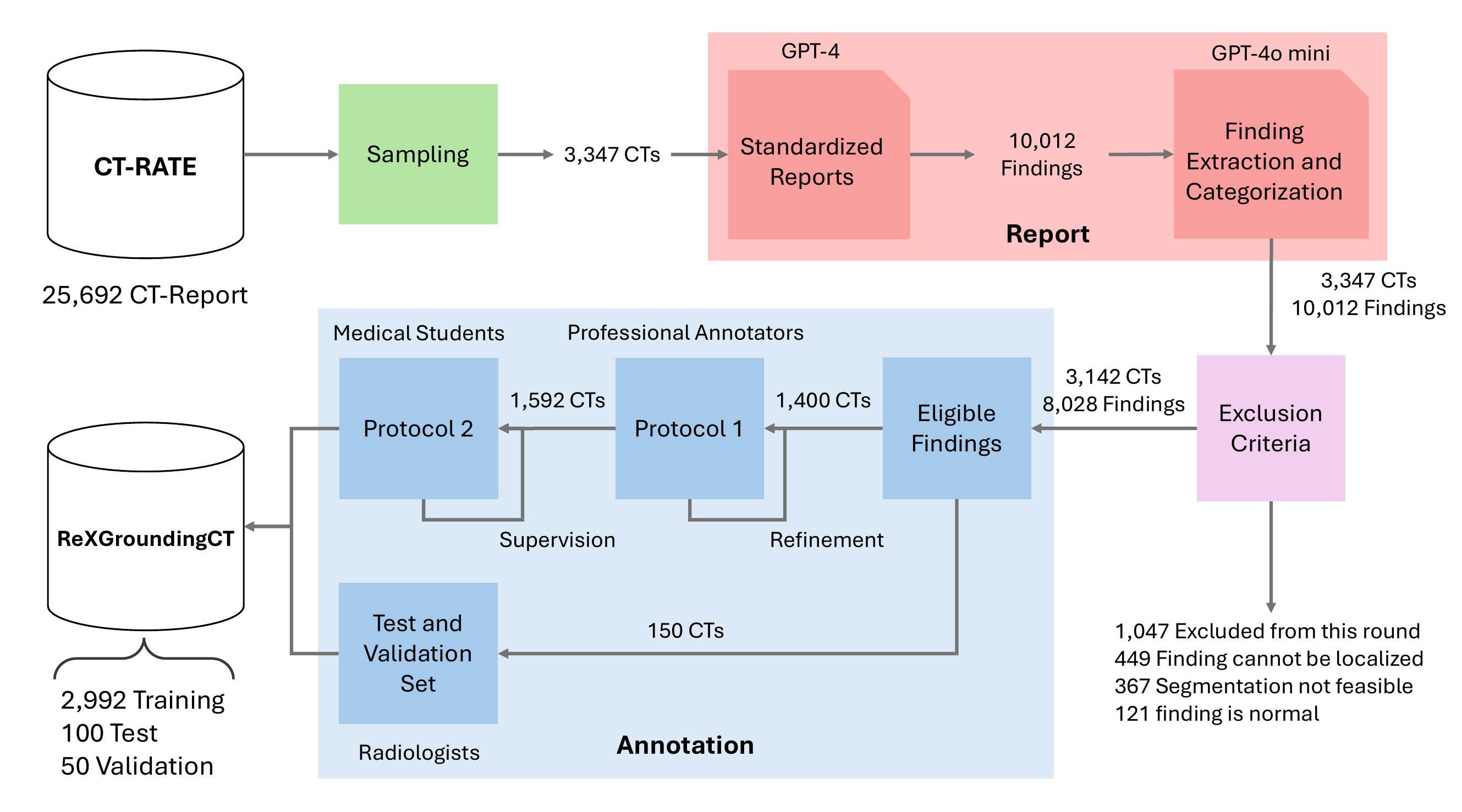}
    \caption{ReXGroundingCT dataset construction pipeline. The training set was annotated with a maximum of three representative instances per finding to manage workload, while the validation and test sets were annotated exhaustively, with all visible instances segmented to provide full ground truth. In protocol 1, all cases were reviewed and refined by board-certified radiologists, ensuring high-quality segmentations. In protocol 2, medical students received structured training and were supervised by senior radiologists, who provided corrections until the students demonstrated sufficient proficiency to annotate independently. The test and validation sets were done exclusively by board-certified radiologists}
    \label{fig:data_pipeline}
\end{figure*}

\begin{figure*}[t!]
    \centering
    \includegraphics[width=1\textwidth]{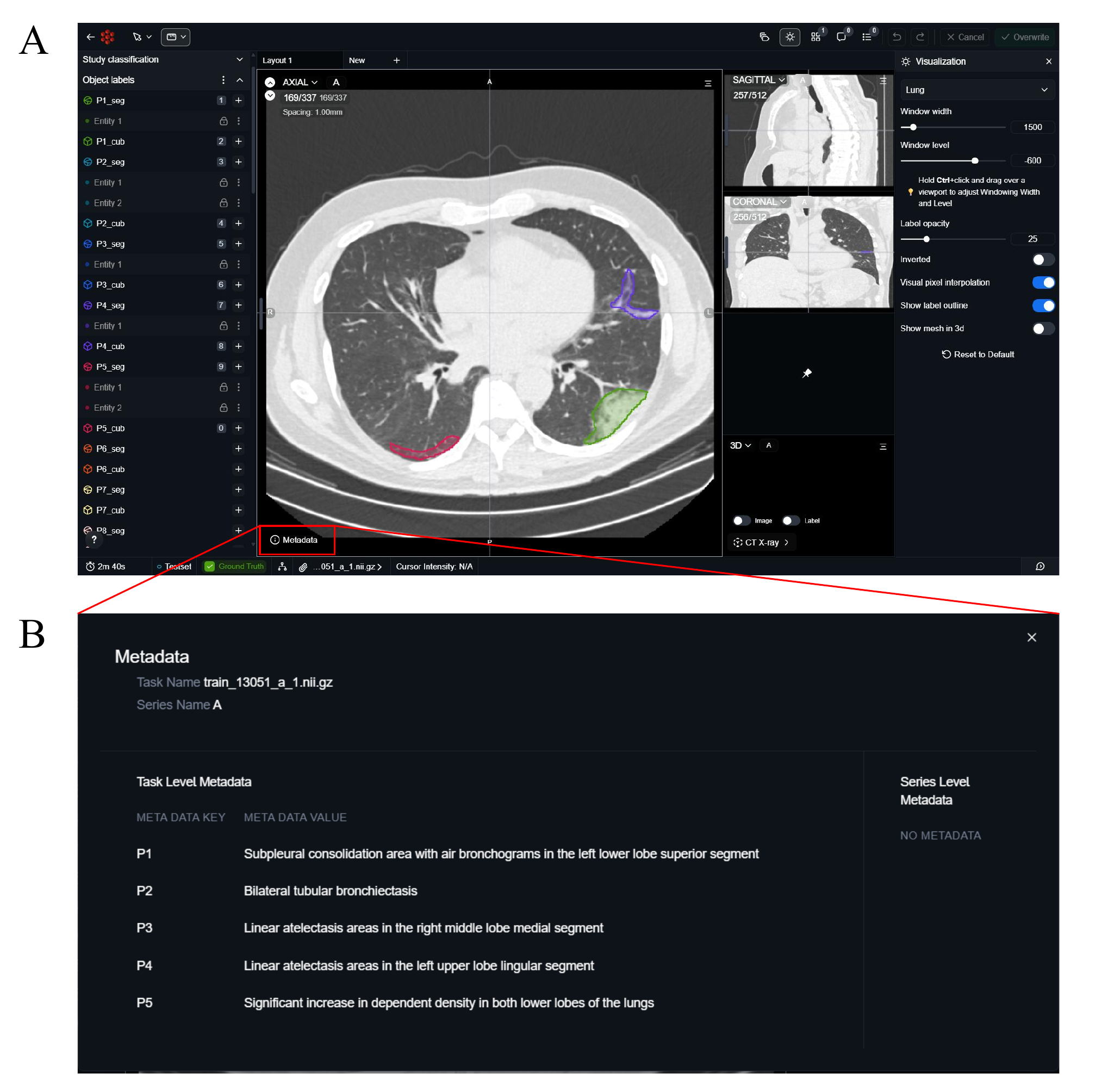}
    \caption{(A) A 3D chest CT scan displayed within the annotation platform, showing multiple pixel-level segmentations corresponding to individual findings (e.g., P1–P5), each color-coded and labeled in the left panel. (B) Associated metadata panel, showing free-text descriptions for each annotated finding. Annotators used this interface to review CT volumes and perform labeling of abnormalities.}
    \label{fig:annotation_platform}
\end{figure*}

All annotations for ReXGroundingCT were performed using a HIPAA-compliant, cloud-based medical imaging annotation platform. Figure \ref{fig:data_pipeline} illustrates the full data construction pipeline, while Figure \ref{fig:annotation_platform} shows the interface of the platform.

\textbf{Annotation Protocols.} The training set included 2,992 chest CT cases, annotated in two different protocols, with quality checks done in both. Using protocol 1, 1,400 cases were annotated by professional annotators, with radiologists refining the cases before accepting them to ensure accuracy. Most annotators have over 1.5 years of medical annotation experience. Radiologists either accepted the segmentations, corrected them directly, or returned them to the annotators for revision when significant discrepancies were identified.

Using the second protocol 2, 1,592 cases were annotated by medical students that are supervised by radiologists. The students underwent dedicated training for this annotation task, during which their performance was monitored and assessed. Incorrect annotations were corrected by supervising radiologists until the students demonstrated sufficient proficiency to perform the annotations independently.

The test and validation sets contain 100 and 50 cases, respectively, and were annotated exclusively by board-certified radiologists, ensuring high-quality and reliable evaluations.

\textbf{Entity Protocol.} The training set followed a different annotation protocol than the validation and test sets. Annotators were instructed to segment no more than three representative instances of a given entity (e.g., up to three nodules), even if more were present. This strategy was used to manage annotation workload. In contrast, the validation and test sets required exhaustive labeling, where annotators segmented all visible instances of each finding to provide complete ground truth for evaluation.

\textbf{Quality control.} Annotation proceeded in two stages: quality control and subsequent segmentation of findings. Across the annotated CT scans, a total of 10,012 distinct findings were initially identified for potential segmentation. During quality control, individual findings were excluded under several specific criteria. A total of 1,047 findings were excluded because they fell outside the scope for this dataset, as only lung and pleura findings were annotated; exclusions in this category included findings such as vascular calcifications or abnormalities in other thoracic or upper abdominal structures. Another 449 findings were excluded because they could not be localized in the CT scan—for example, when a report described a small nodule at a specific location that was not visible to the annotator. An additional 367 findings were excluded because segmentation or bounding was not feasible due to their diffuse nature, including patterns like “decreased aeration in the lung” or extensive “mosaic patterns of attenuation,” which would have required segmenting large portions or the entirety of the lung volume. Lastly, 121 findings were excluded because they represented normal anatomical structures or benign variations, such as subcentimeter lymph nodes considered non-pathological. After applying these exclusion criteria, 8,028 findings remained in the dataset and were included for segmentation and subsequent modeling.

In the segmentation stage, annotators delineated each selected finding within the 3D CT volumes using pixel-level masks. A suite of tools—including brush, polygon, and smart region-growing features—was available to support precise and efficient segmentation of abnormalities. The annotation interface also allowed real-time navigation across slices and adjustment of window levels to enhance visibility during labeling. 

\subsection{Anatomical Chain-of-Thought}

Radiologists naturally follow a systematic spatial reasoning process when interpreting chest CT scans, mentally navigating from broad anatomical regions to precise locations where findings manifest. To enable AI systems to replicate this clinical reasoning pattern, we constructed a chain-of-thought (CoT) dataset that captures the hierarchical anatomical thought process radiologists employ when localizing findings within 3D chest CT volumes.

\begin{figure*}[t!]
    \centering
    \includegraphics[width=1\textwidth]{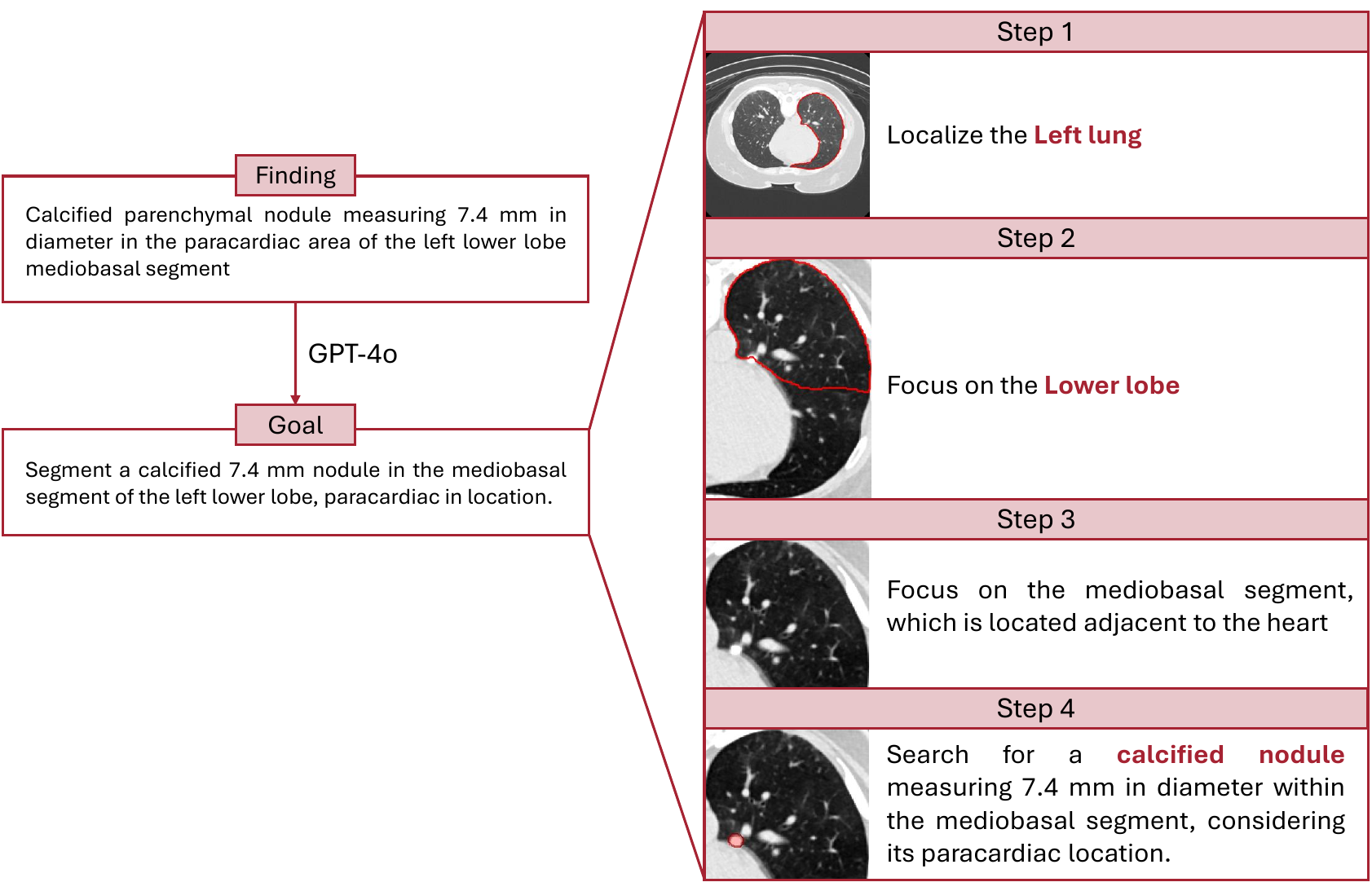}
    \caption{Example of an anatomical chain-of-thought (CoT) reasoning chain. The example illustrates a coarse-to-fine reasoning chain that progressively narrows localization from the whole CT volume to the relevant organ, lobe, and adjacent structures, producing a step-by-step description aligned with the finding text. The lung and lobe boundaries are generated automatically from organ segmentation models \cite{zhang2024radgenome}.}
    \label{fig:cot_example}
\end{figure*}

Motivated by this observation, we built the CoT dataset to complement ReXGroundingCT by explicitly modeling how radiologists reason through anatomical hierarchy when identifying abnormalities. It provides step-by-step hierarchical reasoning sequences that describe how to localize each finding within a 3D chest CT volume. Each reasoning trace begins at the whole-CT level and proceeds through relevant anatomical structures (e.g., lung, lobe, segment) to the region containing the finding, integrating spatial cues such as adjacency to major vessels or pleural surfaces. GPT-4o was used to generate these CoT traces, guided by bounding box coordinates of organs and lobes derived from RadGenome-Chest CT \cite{zhang2024radgenome}. These CoT annotations offer a structured view of spatial reasoning that can support multimodal alignment or serve as supervision for reasoning-augmented segmentation models. A representative CoT example from the dataset is shown in Figure~\ref{fig:cot_example}.

To assess the quality of the CoT dataset, we performed human validation on a randomly selected subset of 150 findings. Each CoT was evaluated by a senior radiologist according to two criteria: (1) \textit{finding alignment}, ensuring that no descriptive details were added or omitted relative to the original finding, and (2) \textit{accuracy of inferred adjacency}, verifying that spatial inferences about neighboring anatomical structures (e.g., “adjacent to pleura,” “near fissure”) were anatomically correct. The validation showed a mean alignment accuracy of 0.92, indicating that the majority of generated reasoning traces were consistent with the original finding descriptions. For the adjacency inference, the accuracy of 0.75, indicating that most inferred spatial relationships were plausible but some CoTs incorrectly assumed pleural or chest wall adjacency when not stated. Only cases in which the model made an adjacency inference were evaluated for this criterion, which was only in around 20\% of the total cases evaluated.

\section{ReXGroundingCT}

The resulting ReXGroundingCT dataset comprises 3,142 chest CT scans and 8,028 segmented findings, representing a diverse range of pulmonary and pleural abnormalities. The average age of patients in the dataset is 42.08 ± 14.52 years, with a sex distribution of 55.9\% male and 44.1\% female. Scans included in ReXGroundingCT were acquired between February 2016 and November 2022, reflecting a broad clinical sampling period across varied patient populations and imaging conditions. All scans are non-contrast chest CT studies, reconstructed with slice thicknesses ranging from 0.50 mm to 5.00 mm, and axial pixel spacing between 0.30 mm and 0.98 mm. Most volumes have matrix sizes of 512 × 512 (92.6\%), though higher resolutions such as 768 × 768 (7.2\%) and 1024 × 1024 (0.2\%) are also present. The number of slices per volume varies from 104 to 1005, with a mean of approximately 351 slices.

In total, the dataset contains 8,028 segmented findings, with 16,301 separate entities, distributed across both focal and non-focal abnormalities. Approximately 79\% of findings are focal lesions, such as nodules or localized consolidations, while the remaining 21\% represent diffuse or infiltrative patterns. Each finding is associated with a precise 3D segmentation mask, enabling detailed spatial analysis.
ReXGroundingCT’s findings span 14 categories (Figure \ref{fig:annotation_example}, B), of which 8 are focal and 6 are non-focal, encompassing diverse pulmonary and pleural abnormalities. The most common focal category is “Pulmonary Nodules/Masses”, accounting for 28.9\% of focal findings, while the most common non-focal category is “Emphysema”, representing 27.7\% of non-focal findings.

In the training set, each finding contains an average of 1.95 entities due to the annotation protocol limiting segmentation to at most three instances per finding. In contrast, scans in the test and validation sets combined contain an average of 3.80 finding entities, reflecting a more comprehensive annotation of all visible abnormalities (see Figure \ref{fig:entities_dist_per_category}).

\section{Results}

We evaluated several state-of-the-art text-prompted 3D medical segmentation models, including BiomedParsev2 \cite{zhao2025foundation}, SAT \cite{zhao2025large}, and SegVol \cite{du2024segvol}, on the ReXGroundingCT test set. As shown on the public leaderboard hosted at \href{https://rexrank.ai/ReXGroundingCT/}{ReXRank} \cite{zhang2025rexrank}, all existing models perform poorly in grounding free-text findings to their corresponding 3D regions, underscoring the difficulty of this task due to the absence of prior datasets that link narrative CT findings to 3D annotations. All off-the-shelf models struggled to localize free-text findings in 3D space, achieving near-zero performance across metrics.

We also show that fine-tuning on ReXGroundingCT substantially improved segmentation accuracy. The fine-tuned SAT model (SAT-FT) achieved a Global Dice of 0.209, Global HIT Rate of 0.473, representing a significant improvement over the baselines. This demonstrates that ReXGroundingCT provides the necessary supervision signal to adapt existing architectures to the challenging task of free-text 3D grounding. Yet, substantial progress still to be made in terms of model architecture, data scale, and representation learning to achieve clinically meaningful performance for this task.

To facilitate continued progress on this task, an active leaderboard is maintained through \href{https://rexrank.ai/ReXGroundingCT/}{ReXRank}, allowing the community to benchmark new models and track performance.

\section{Limitations}

While ReXGroundingCT represents a significant advance in linking free-text radiology findings to 3D segmentations, several limitations should be considered.

First, annotations in the training set were performed by professional annotators and medical students, rather than exclusively by board-certified radiologists. This introduces potential variability in annotation quality. However, to mitigate this issue, all annotations underwent structured quality control processes. In Protocol 1, radiologists reviewed and directly corrected annotations or returned them for revision. In Protocol 2, medical students were trained under the supervision of senior radiologists, who monitored their progress and corrected all errors until the students demonstrated sufficient proficiency.

Second, in the training set, annotators were instructed to segment no more than three representative instances per finding. This partial labeling strategy, while essential for managing the annotation workload across thousands of volumes, limits the completeness of spatial annotations. Nonetheless, this limitation can be addressed in model development using weakly supervised or semi-supervised learning techniques that can leverage incomplete annotations to infer full spatial representations \cite{wolny2022sparse, gao2022segmentation}.

Finally, the dataset focuses exclusively on lung and pleural findings, with non-pulmonary findings excluded during quality control. This restricts the dataset’s applicability to other thoracic or abdominal abnormalities. Expanding the annotation scope to include these findings could improve the dataset’s generalizability for comprehensive report grounding tasks.

\section{Discussion}

ReXGroundingCT is the first publicly available manually annotated dataset to link free text radiology findings with precise pixel-level segmentations in 3D chest CT scans. Unlike prior datasets, which rely on categorical labels or structured annotations, ReXGroundingCT captures the full variability of clinical language through sentence-level grounding to volumetric imaging.

ReXGroundingCT supports two core tasks central to advancing multimodal medical AI: (1) \textit{finding grounding}, which involves localizing the spatial extent of a specific free-text finding in a 3D CT scan, and (2) \textit{grounded report generation}, where a model must generate descriptive radiology reports with spatial references aligned to pixel-level segmentations \cite{bannur2024maira}. These tasks move beyond traditional classification or bounding-box localization by requiring fine-grained spatial reasoning and natural language understanding within volumetric data.

\begin{figure*}[t!]
    \centering
    \includegraphics[width=1\textwidth]{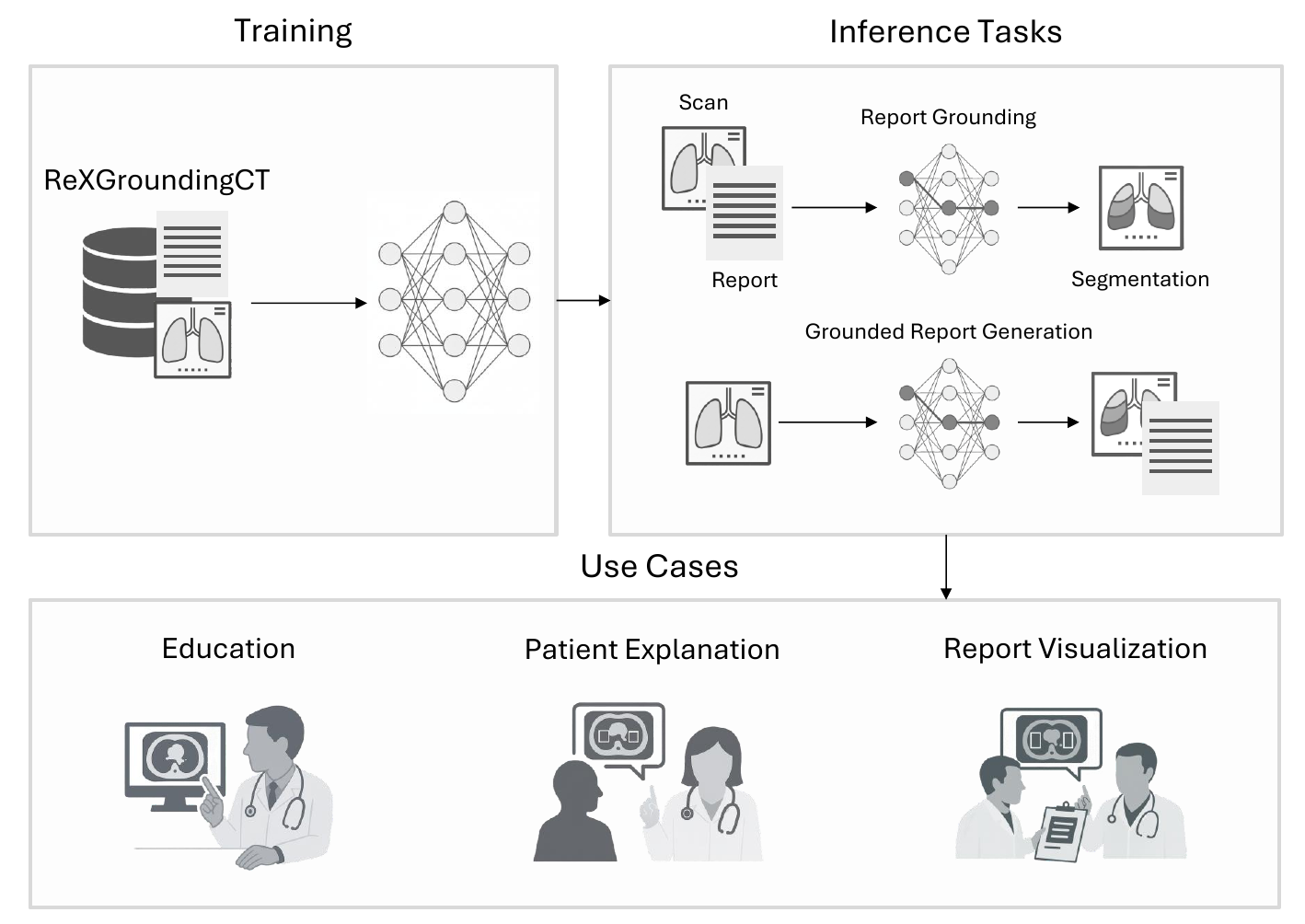}
    \caption{Potential downstream applications of ReXGroundingCT. Training on the ReXGroundingCT dataset can enable models for two key tasks: (1) \textit{report grounding}, which localizes findings described in a new patient's report, and (2) \textit{grounded report generation}, which produces spatially grounded reports directly from CT scans. These capabilities support three key downstream use cases: (1) radiologist education, exemplified by \textit{RadGame}~\cite{baharoon2025radgame}, an AI-powered platform designed to improve localization and report-writing capabilities of future radiologists; (2) patient explanation, demonstrated by \textit{ReXplain}~\cite{luo2025rexplain}, which visually communicates findings on patient scans to enhance understanding; and (3) report visualization, which augments radiology reports with spatially grounded imaging overlays to improve communication between radiologists and referring physicians.}

    \label{fig:future_figure}
\end{figure*}

Linking free-text radiology reports to precise 3D segmentations also has direct clinical relevance. Such tools could reduce the time and effort required for radiologists and clinicians to correlate narrative reports with imaging findings, facilitate clearer communication with referring physicians, and potentially reduce unnecessary consultations \cite{iyer2010added}. For patients, visual context directly linked to report language may improve comprehension and engagement with their health information \cite{luo2025rexplain}. Radiology trainees stand to benefit from systems that visually connect descriptive findings to anatomical regions, supporting the development of spatial reasoning skills essential for diagnostic accuracy \cite{karstens2024possible}. These use cases and tasks are illustrated in Figure \ref{fig:future_figure}.

Although several chest CT datasets have been developed to support disease classification \cite{afshar2021covid} and lesion segmentation tasks \cite{armato2011lung, antonelli2022medical, bilic2023liver}, they typically rely on structured metadata or limited label taxonomies. These constraints limit their ability to support open-ended language inputs or nuanced radiological descriptions. More similar to our work, PadChest-GR introduced sentence-level grounding to address this gap but is limited to chest X-rays and bounding-box annotations \cite{de2025padchest}. OminiAbnorm-CT \cite{zhao2025rethinking} introduces a large-scale dataset for grounding and describing abnormalities in multi-plane CT across the whole body, but is limited to 2D key slices from Radiopaedia and requires a paid license for use in machine learning applications, which restricts accessibility. RadGenome-Chest CT \cite{zhang2024radgenome} provides region-grounded supervision using automated organ-level segmentations and sentence-region matching for chest CT, but does not focus on disease-level findings. RadGPT \cite{bassi2025radgpt} introduces AbdomenAtlas 3.0, a large-scale abdominal CT dataset where all structured and narrative radiology reports are automatically generated from per-voxel tumor and organ segmentations using a deterministic pipeline and large language models. MedTrinity-25M \cite{xiemedtrinity} provides over 25 million multigranular image–ROI–description triplets across ten imaging modalities, generated automatically using expert models and retrieval-augmented multimodal language models. ReXGroundingCT advances this line of work by providing expert-monitored, disease-level sentence grounding in volumetric chest CT with dense, pixel-level segmentations that are manually annotated.

\section{Conclusion}

ReXGroundingCT introduces the first manually annotated publicly available dataset that links free-text radiology findings to expert-monitored 3D segmentations in chest CT scans. By capturing the full expressiveness of clinical language and grounding it in volumetric space, the dataset addresses a critical gap in the development of grounded radiology report generation systems. Through a structured pipeline involving standardized report rewriting, high-fidelity finding extraction and categorization, and quality-assured pixel-level annotation, ReXGroundingCT enables new research directions in sentence-level grounding. We hope this resource will serve as a foundation for advancing explainable and anatomically grounded AI systems in medical imaging.

\section{Acknowledgement}

This research was supported by a grant of the Boston-Korea Innovative Research Project through the Korea Health Industry Development Institute (KHIDI), funded by the Ministry of Health \& Welfare, Republic of Korea (Grant number: RS-2024-00403047).

This research was also supported by Harvard Medical School Dean’s Innovation Award for Accelerating Foundation Model Research

We also thank Dr. Abdulrhman Aljouie for facilitating the international collaboration with King Abdullah Specialized Children’s Hospital and King Abdulaziz Medical City in Riyadh, Saudi Arabia.

\bibliographystyle{plain}
\bibliography{main}

\begin{thebibliography}{10}

\bibitem{abgrall2024should}
Gw{\'e}nol{\'e} Abgrall, Andre~L Holder, Zaineb Chelly~Dagdia, Karine Zeitouni, and Xavier Monnet.
\newblock Should ai models be explainable to clinicians?
\newblock {\em Critical Care}, 28(1):301, 2024.

\bibitem{achiam2023gpt}
Josh Achiam, Steven Adler, Sandhini Agarwal, Lama Ahmad, Ilge Akkaya, Florencia~Leoni Aleman, Diogo Almeida, Janko Altenschmidt, Sam Altman, Shyamal Anadkat, et~al.
\newblock Gpt-4 technical report.
\newblock {\em arXiv preprint arXiv:2303.08774}, 2023.

\bibitem{afshar2021covid}
Parnian Afshar, Shahin Heidarian, Nastaran Enshaei, Farnoosh Naderkhani, Moezedin~Javad Rafiee, Anastasia Oikonomou, Faranak~Babaki Fard, Kaveh Samimi, Konstantinos~N Plataniotis, and Arash Mohammadi.
\newblock Covid-ct-md, covid-19 computed tomography scan dataset applicable in machine learning and deep learning.
\newblock {\em Scientific Data}, 8(1):121, 2021.

\bibitem{antonelli2022medical}
Michela Antonelli, Annika Reinke, Spyridon Bakas, Keyvan Farahani, Annette Kopp-Schneider, Bennett~A Landman, Geert Litjens, Bjoern Menze, Olaf Ronneberger, Ronald~M Summers, et~al.
\newblock The medical segmentation decathlon.
\newblock {\em Nature communications}, 13(1):4128, 2022.

\bibitem{armato2011lung}
Samuel~G Armato~III, Geoffrey McLennan, Luc Bidaut, Michael~F McNitt-Gray, Charles~R Meyer, Anthony~P Reeves, Binsheng Zhao, Denise~R Aberle, Claudia~I Henschke, Eric~A Hoffman, et~al.
\newblock The lung image database consortium (lidc) and image database resource initiative (idri): a completed reference database of lung nodules on ct scans.
\newblock {\em Medical physics}, 38(2):915--931, 2011.

\bibitem{baharoon2025radgame}
Mohammed Baharoon, Siavash Raissi, John~S Jun, Thibault Heintz, Mahmoud Alabbad, Ali Alburkani, Sung~Eun Kim, Kent Kleinschmidt, Abdulrahman~O Alhumaydhi, Mohannad Mohammed~G Alghamdi, et~al.
\newblock Radgame: An ai-powered platform for radiology education.
\newblock {\em arXiv preprint arXiv:2509.13270}, 2025.

\bibitem{bannur2024maira}
Shruthi Bannur, Kenza Bouzid, Daniel~C Castro, Anton Schwaighofer, Anja Thieme, Sam Bond-Taylor, Maximilian Ilse, Fernando P{\'e}rez-Garc{\'\i}a, Valentina Salvatelli, Harshita Sharma, et~al.
\newblock Maira-2: Grounded radiology report generation.
\newblock {\em arXiv preprint arXiv:2406.04449}, 2024.

\bibitem{bassi2025radgpt}
Pedro~RAS Bassi, Mehmet~Can Yavuz, Kang Wang, Xiaoxi Chen, Wenxuan Li, Sergio Decherchi, Andrea Cavalli, Yang Yang, Alan Yuille, and Zongwei Zhou.
\newblock Radgpt: Constructing 3d image-text tumor datasets.
\newblock {\em arXiv preprint arXiv:2501.04678}, 2025.

\bibitem{bilic2023liver}
Patrick Bilic, Patrick Christ, Hongwei~Bran Li, Eugene Vorontsov, Avi Ben-Cohen, Georgios Kaissis, Adi Szeskin, Colin Jacobs, Gabriel Efrain~Humpire Mamani, Gabriel Chartrand, et~al.
\newblock The liver tumor segmentation benchmark (lits).
\newblock {\em Medical image analysis}, 84:102680, 2023.

\bibitem{de2025padchest}
Daniel~Coelho de~Castro, Aurelia Bustos, Shruthi Bannur, Stephanie~L Hyland, Kenza Bouzid, Maria~Teodora Wetscherek, Maria~Dolores S{\'a}nchez-Valverde, Lara Jaques-P{\'e}rez, Lourdes P{\'e}rez-Rodr{\'\i}guez, Kenji Takeda, et~al.
\newblock Padchest-gr: A bilingual chest x-ray dataset for grounded radiology report generation.
\newblock {\em NEJM AI}, 2(7):AIdbp2401120, 2025.

\bibitem{de2025uls23}
MJJ de~Grauw, E~Th Scholten, Ewoud~J Smit, Matthieu~JCM Rutten, M~Prokop, B~van Ginneken, and Alessa Hering.
\newblock The uls23 challenge: A baseline model and benchmark dataset for 3d universal lesion segmentation in computed tomography.
\newblock {\em Medical Image Analysis}, 102:103525, 2025.

\bibitem{du2024segvol}
Yuxin Du, Fan Bai, Tiejun Huang, and Bo~Zhao.
\newblock Segvol: Universal and interactive volumetric medical image segmentation.
\newblock {\em Advances in Neural Information Processing Systems}, 37:110746--110783, 2024.

\bibitem{fahrner2025generative}
L~John Fahrner, Emma Chen, Eric Topol, and Pranav Rajpurkar.
\newblock The generative era of medical ai.
\newblock {\em Cell}, 188(14):3648--3660, 2025.

\bibitem{gao2022segmentation}
Feng Gao, Minhao Hu, Min-Er Zhong, Shixiang Feng, Xuwei Tian, Xiaochun Meng, Zeping Huang, Minyi Lv, Tao Song, Xiaofan Zhang, et~al.
\newblock Segmentation only uses sparse annotations: Unified weakly and semi-supervised learning in medical images.
\newblock {\em Medical Image Analysis}, 80:102515, 2022.

\bibitem{hamamci2024developing}
Ibrahim~Ethem Hamamci, Sezgin Er, Chenyu Wang, Furkan Almas, Ayse~Gulnihan Simsek, Sevval~Nil Esirgun, Irem Doga, Omer~Faruk Durugol, Weicheng Dai, Murong Xu, et~al.
\newblock Developing generalist foundation models from a multimodal dataset for 3d computed tomography.
\newblock {\em arXiv preprint arXiv:2403.17834}, 2024.

\bibitem{iyer2010added}
Veena~R Iyer, Peter~F Hahn, Lawrence~S Blaszkowsky, Sarah~P Thayer, Elkan~F Halpern, and Mukesh~G Harisinghani.
\newblock Added value of selected images embedded into radiology reports to referring clinicians.
\newblock {\em Journal of the American College of Radiology}, 7(3):205--210, 2010.

\bibitem{karstens2024possible}
Jort Karstens, Lucas Boer, Thomas Maal, and Dylan Henssen.
\newblock The possible impact of segmenting radiological images on students’ spatial abilities.
\newblock {\em Discover Education}, 3(1):83, 2024.

\bibitem{luo2025rexplain}
Luyang Luo, Jenanan Vairavamurthy, Xiaoman Zhang, Abhinav Kumar, Ramon~R Ter-Oganesyan, Stuart~T Schroff, Dan Shilo, Rydhwana Hossain, Mike Moritz, and Pranav Rajpurkar.
\newblock Rexplain: Translating radiology into patient-friendly video reports.
\newblock In {\em AAAI Bridge Program on AI for Medicine and Healthcare}, pages 109--120. PMLR, 2025.

\bibitem{rao2025multimodal}
Vishwanatha~M Rao, Michael Hla, Michael Moor, Subathra Adithan, Stephen Kwak, Eric~J Topol, and Pranav Rajpurkar.
\newblock Multimodal generative ai for medical image interpretation.
\newblock {\em Nature}, 639(8056):888--896, 2025.

\bibitem{sadeghi2024review}
Zahra Sadeghi, Roohallah Alizadehsani, Mehmet~Akif Cifci, Samina Kausar, Rizwan Rehman, Priyakshi Mahanta, Pranjal~Kumar Bora, Ammar Almasri, Rami~S Alkhawaldeh, Sadiq Hussain, et~al.
\newblock A review of explainable artificial intelligence in healthcare.
\newblock {\em Computers and Electrical Engineering}, 118:109370, 2024.

\bibitem{silcox2024potential}
Christina Silcox, Eyal Zimlichmann, Katie Huber, Neil Rowen, Robert Saunders, Mark McClellan, Charles~N Kahn~III, Claudia~A Salzberg, and David~W Bates.
\newblock The potential for artificial intelligence to transform healthcare: perspectives from international health leaders.
\newblock {\em NPJ Digital Medicine}, 7(1):88, 2024.

\bibitem{wasserthal2023totalsegmentator}
Jakob Wasserthal, Hanns-Christian Breit, Manfred~T Meyer, Maurice Pradella, Daniel Hinck, Alexander~W Sauter, Tobias Heye, Daniel~T Boll, Joshy Cyriac, Shan Yang, et~al.
\newblock Totalsegmentator: robust segmentation of 104 anatomic structures in ct images.
\newblock {\em Radiology: Artificial Intelligence}, 5(5):e230024, 2023.

\bibitem{wolny2022sparse}
Adrian Wolny, Qin Yu, Constantin Pape, and Anna Kreshuk.
\newblock Sparse object-level supervision for instance segmentation with pixel embeddings.
\newblock In {\em Proceedings of the IEEE/CVF Conference on Computer Vision and Pattern Recognition}, pages 4402--4411, 2022.

\bibitem{xiemedtrinity}
Yunfei Xie, Ce~Zhou, Lang Gao, Juncheng Wu, Xianhang Li, Hong-Yu Zhou, Sheng Liu, Lei Xing, James Zou, Cihang Xie, et~al.
\newblock Medtrinity-25m: A large-scale multimodal dataset with multigranular annotations for medicine.
\newblock In {\em The Thirteenth International Conference on Learning Representations}.

\bibitem{zhang2024radgenome}
Xiaoman Zhang, Chaoyi Wu, Ziheng Zhao, Jiayu Lei, Weiwei Tian, Ya~Zhang, Weidi Xie, and Yanfeng Wang.
\newblock Development of a large-scale grounded vision language dataset for chest ct analysis.
\newblock {\em Scientific Data}, 12(1):1636, 2025.

\bibitem{zhang2025rexrank}
Xiaoman Zhang, Hong-Yu Zhou, Xiaoli Yang, Oishi Banerjee, Juli{\'a}n~N Acosta, Josh Miller, Ouwen Huang, and Pranav Rajpurkar.
\newblock Rexrank: A public leaderboard for ai-powered radiology report generation.
\newblock In {\em AAAI Bridge Program on AI for Medicine and Healthcare}, pages 90--99. PMLR, 2025.

\bibitem{zhao2025foundation}
Theodore Zhao, Yu~Gu, Jianwei Yang, Naoto Usuyama, Ho~Hin Lee, Sid Kiblawi, Tristan Naumann, Jianfeng Gao, Angela Crabtree, Jacob Abel, et~al.
\newblock A foundation model for joint segmentation, detection and recognition of biomedical objects across nine modalities.
\newblock {\em Nature methods}, 22(1):166--176, 2025.

\bibitem{zhao2025rethinking}
Ziheng Zhao, Lisong Dai, Ya~Zhang, Yanfeng Wang, and Weidi Xie.
\newblock Rethinking whole-body ct image interpretation: An abnormality-centric approach.
\newblock {\em arXiv preprint arXiv:2506.03238}, 2025.

\bibitem{zhao2025large}
Ziheng Zhao, Yao Zhang, Chaoyi Wu, Xiaoman Zhang, Xiao Zhou, Ya~Zhang, Yanfeng Wang, and Weidi Xie.
\newblock Large-vocabulary segmentation for medical images with text prompts.
\newblock {\em NPJ Digital Medicine}, 8(1):566, 2025.

\end{thebibliography}

\newpage

\appendix
\renewcommand\thefigure{\thesection.\arabic{figure}}
\setcounter{figure}{0}
\onecolumn

\section{Number of Entities for Each Finding Category}

\begin{figure*}[htbp]
    \centering
    \includegraphics[width=1\textwidth]{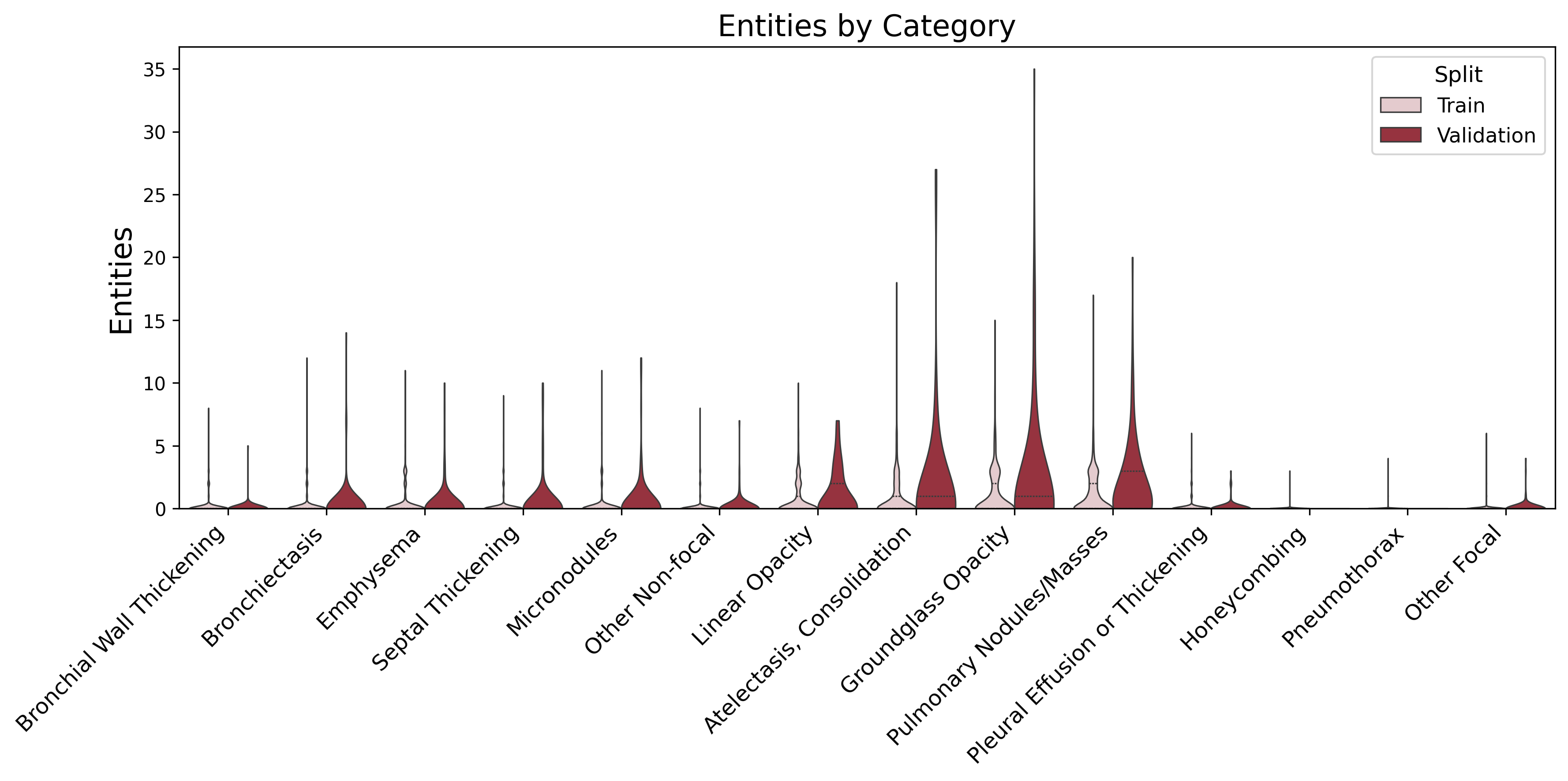}
    \caption{Distribution of segmented entities per finding category. Validation distributions show greater mass at higher entity counts (e.g., >3), reflecting the exhaustive labeling protocol in which all visible instances were segmented. In contrast, the training set was limited to at most three representative instances per finding to reduce annotation workload.}
    \label{fig:entities_dist_per_category}
\end{figure*}

\newpage

\section{Relative size of Finding Categories}
\setcounter{figure}{0}

\begin{figure*}[htbp]
    \centering
    \includegraphics[width=1\textwidth]{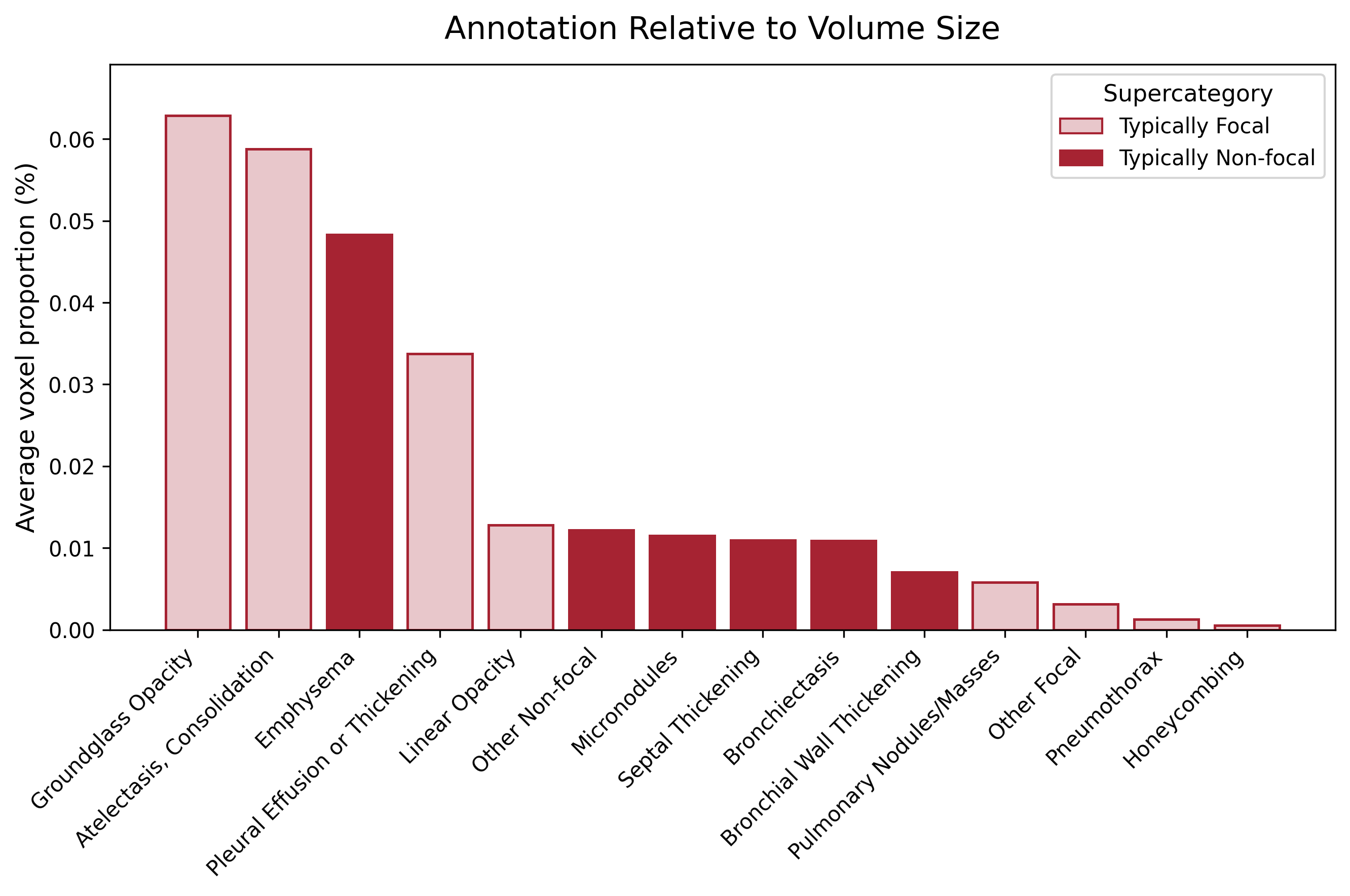}
    \caption{Average annotation ground truth size with respect to volume size per finding category.}
    \label{fig:relative_size_per_category}
\end{figure*}

\newpage

\setcounter{figure}{0}
\section{Examples from the Dataset}

\begin{figure*}[htbp]
    \centering
    \includegraphics[width=0.8\textwidth]{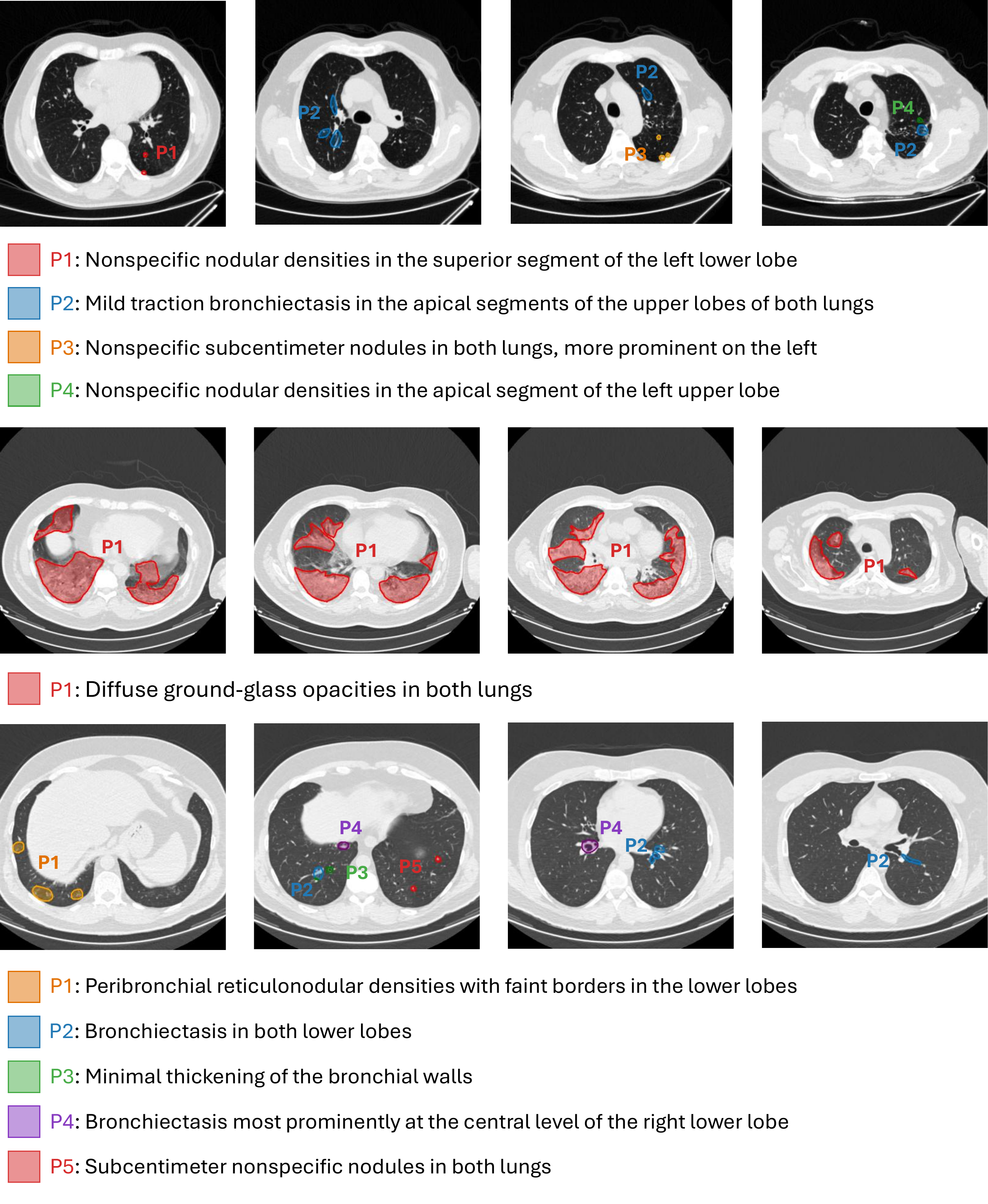}
    \caption{Additional Examples from the dataset.}
    \label{}
\end{figure*}

\newpage

\section{Hierarchical Tree Finding Categories}
\label{fig:finding_categories} 
\dirtree{%
.1 1) Typically non-focal lung/airway/pleural abnormalities.
.2 1a) Bronchial wall thickening.
.2 1b) Bronchiectasis.
.2 1c) Emphysema (including Centrilobular, Paraseptal, Bullous).
.2 1d) Septal thickening (including Interlobular, Reticulation).
.2 1e) Micronodules (including Centrilobular, Tree-in-bud, Perilymphatic).
.2 1f) Other.
.1 2) Typically focal lung/airway/pleural opacities.
.2 2a) Linear (including subsegmental atelectasis, scarring, fibrosis).
.2 2b) Atelectasis, consolidation.
.2 2c) Groundglass opacity.
.2 2d) Pulmonary nodules/masses.
.2 2e) Pleural effusion or thickening.
.2 2f) Honeycombing.
.2 2g) Pneumothorax.
.2 2h) Other.
.1 3) Non-pulmonary lesions.
.2 3a) Lymphadenopathy lesions.
.2 3b) Liver lesions.
.2 3c) Gallbladder lesions.
.2 3d) Renal/kidneys, collecting system, and ureters lesions.
.2 3e) Spleen lesions.
.2 3f) Adrenal lesions.
.2 3g) Pancreas lesions.
.2 3h) Thyroid lesions.
.2 3i) Skin/subcutaneous lesions.
.2 3j) Bone/Osseous structures lesions.
.2 3k) Other lesions.
.1 4) Bones (non-lesion).
.2 4a) Fractures.
.2 4b) Degenerative joint disease, degenerative disc disease, arthritis.
.2 4c) Spinal curvature abnormalities: kyphosis, scoliosis.
.2 4d) Other.
.1 5) Stones/organ calcifications (non-lesion).
.2 5a) Nephroliths, choleliths.
.2 5b) Granulomas.
.2 5c) Other.
.1 6) Hollow viscera abnormalities.
.2 6a) Hiatus hernia.
.2 6b) Wall thickening.
.2 6c) Dilated.
.2 6d) Diverticulum (including diverticulosis).
.2 6e) Other.
.1 7) Skin/subcutaneous (non-lesion).
.2 7a) Skin thickening.
.2 7b) Stranding.
.2 7c) Abdominal wall hernia: ventral, umbilical, inguinal.
.2 7d) Gynecomastia.
.2 7e) Other.
.1 8) Cardiovascular.
.2 8a) Atherosclerosis (including coronary, non-coronary).
.2 8b) Vessel aneurysm, ectasia, enlargement.
.2 8c) Vessel occlusion or stenosis.
.2 8d) Cardiac chamber enlargement.
.2 8e) Valvular calcification.
.2 8f) Pericardial effusion.
.2 8g) Other.
.1 9) Body composition.
.2 9a) Visceral fat.
.2 9b) Superficial subcutaneous fat.
.2 9c) Skeletal muscle.
.2 9d) Osteoporosis.
.2 9e) Hepatic steatosis.
.2 9f) Other.
.1 10) Diffuse/whole organ.
.2 10a) Organomegaly (including splenomegaly, multinodular goiter, thyromegaly, lung hyperinflation).
.2 10b) Atrophy.
.2 10c) Other.
.1 11) Device.
.2 11a) Elongated (including catheter, pacemaker/defibrillator, spinal stimulator).
.2 11b) Surgical clips.
.2 11c) Other.
.1 12) Other.
}

\end{document}